\documentclass[aps,prd,twocolumn,superscriptaddress,showpacs,preprintnumbers]{revtex4}

\usepackage{array}
\usepackage{dcolumn}
\usepackage{graphics}
\usepackage{hyperref}

\usepackage{color}
\definecolor{IITred}{rgb}{0.5,0.05,0.05}
\input boxedeps
\SetepsfEPSFSpecial
\HideDisplacementBoxes

\newcommand{\slj}[3]{\mbox{$^{#1}${\ifcase#2\or S\or 
         P\or D\or F\or G\fi}$_{#3}$}}
\newcommand{\sLj}[3]{{}^{#1}\!#2_{#3}}
	 
\newcommand{\ev}{\hbox{ eV}}
\newcommand{\kev}{\hbox{ keV}}
\newcommand{\mev}{\hbox{ MeV}}
\newcommand{\gev}{\hbox{ GeV}}

\newcommand{\etal}{{\em et al.}}
\newcommand{\ccc}{C$^{3}$}

\newcommand{\zcc}{\mathcal{Z}_{c\bar{c}}}

\newcommand{\cfrac}[2]{\textstyle{\frac{#1}{#2}}}

\newcommand{\jpsi}{\ensuremath{J\!/\!\psi}}

\def\Ket#1{\left| #1\right\rangle}

\begin{document}

\preprint{FERMILAB--Pub--05/380--T--rev}
\preprint{BUHEP-05-18}

\title{New states above charm threshold}


\author{Estia J. Eichten}
\email[E-mail: ]{eichten@fnal.gov}
\affiliation{Theoretical Physics Department\\ Fermi National 
Accelerator Laboratory\\ P.O.\ Box 500, Batavia, IL 60510}
\author{Kenneth Lane}
\email[E-mail: ]{lane@bu.edu}
\affiliation{Department of Physics, Boston
University\\ 590 Commonwealth Avenue, Boston, MA 02215}
\author{Chris Quigg}
\email[E-mail: ]{quigg@fnal.gov}
\affiliation{Theoretical Physics Department\\ Fermi National 
Accelerator Laboratory\\ P.O.\ Box 500, Batavia, IL 60510}



\begin{abstract}
We revise and extend expectations for the properties of charmonium 
states that lie above charm threshold, in light of new experimental 
information. We refine the Cornell coupled-channel model for the 
coupling of $c\bar{c}$ levels to two-meson states, defining resonance 
masses and widths by pole positions in the complex energy plane, and
suggest new targets for experiment.
\end{abstract}

\pacs{14.40.Gx,13.25.Gv,14.40.Lb}

\maketitle

\section{Introduction\label{sec:intro}}

In the short time since the Belle Collaboration reported sighting the
``missing'' charmonium state $\eta_{c}^{\prime}(2\slj{1}{1}{0})$ in
exclusive $B \to K K_{S} K^{-}\pi^{+}$ decays~\cite{Choi:2002na}, new
states associated with charmonium have appeared in great profusion.
Apart from the long-sought $h_{c}(1\slj{1}{2}{1})$, recently observed
by the CLEO Collaboration in the decay $\psi(2\mathrm{S}) \to
\pi^{0}h_{c} \to (\gamma\gamma)(\gamma\eta_{c})$~\cite{Rosner:2005ry,Rubin:2005px},
all the newly observed states lie near or above charm threshold. The 
identification of levels above the $D\bar{D}$ flavor threshold has renewed the 
importance of theoretical studies that couple the
$c\bar{c}$ spectrum to charmed-meson--anticharmed-meson channels.

Motivated by the Belle Collaboration's discovery~\cite{Choi:2003ue} of
the narrow state $X(3872) \to \pi^{+}\pi^{-}\jpsi$, confirmed in short
order by the CDF~\cite{Acosta:2003zx}, D\O~\cite{Abazov:2004kp}, and
BaBar~\cite{Aubert:2004ns} Experiments, we explored the influence of
open-charm channels on charmonium properties, and profiled the
1\slj{3}{3}{2}, 1\slj{3}{3}{3}, and 2\slj{1}{2}{1} charmonium
candidates for $X(3872)$~\cite{Eichten:2004uh}.  We reaffirmed the
single-channel potential-model
expectation~\cite{Eichten:2002qv,Barnes:2003vb} that the favored
1\slj{3}{3}{2} and 1\slj{3}{3}{3} candidates both should have prominent
radiative decays, and noted that the 1\slj{3}{3}{2} might be visible in
the $D^{0}\bar{D}^{*0}$ channel, while the dominant decay of the
1\slj{3}{3}{3} state should be into $D\bar{D}$.  We proposed that
additional discrete charmonium levels were ripe for discovery as narrow
resonances of charmed and anticharmed mesons.  Barnes, Godfrey, and
Swanson have reached similar conclusions using a different model for
the coupling to open charm~\cite{Barnes:2005pb}.

Experiments have enriched our knowledge of $X(3872)$, but its precise
nature remains open to debate.  On current evidence, $X(3872)$ is
likely to be a $J^{PC}=1^{++}$ state.  If so, the surviving
2\slj{3}{2}{1} charmonium candidate is problematic, as we discuss in
\S\ref{subsubsec:x23p1}.  Not every state associated with charmonium need be
identified as a $c\bar{c}$ state; we will mention below what some of
the interlopers might be.  An important goal of predicting the
properties of charmonium states above flavor threshold is to tag new
states that do not fit the template, and so might represent new
spectroscopies.

Four other new particles require confirmation, for each has 
been seen only in a single experiment so far.
Belle~\cite{Abe:2004zs} has reported $Y(3943
\pm 11 \pm 13)$ in the decay $B \to K\omega\jpsi$; it is a relatively broad
state, with total width $\Gamma = 87 \pm 22\pm 26\mev$.  Belle also reports
the state $X(3943 \pm 9)$, seen in $e^{+}e^{-} \to \jpsi +
X$~\cite{Abe:2005hd}.  It is observed to decay into $D\bar{D}^{*}$, but
not $D\bar{D}$, which suggests an unnatural parity assignment.  The
total width is $\Gamma < 52\mev$.  

Belle has observed a narrow ($\Gamma \approx 20\mev$) state in
$\gamma\gamma \to D\bar{D}$ that they call $Z(3929 \pm 5 \pm
2)$~\cite{Uehara:2005qd}.  The production and decay characteristics are
consistent with a $2^{++}$ assignment, and this state is a plausible
$\chi_{c2}^{\prime}$ (2\slj{3}{2}{2}) candidate.  The most recent
addition to the collection is $Y(4260)$, a broad ($\approx 88\mev$)
$1^{--}$ level seen by BaBar in $e^{+}e^{-} \to \gamma
\pi^{+}\pi^{-}\,\jpsi$~\cite{Aubert:2005rm}, with supporting evidence
from $B \to K^{-}\jpsi\pi\pi$~\cite{Aubert:2005zh}.

In this Article, we fix the masses of the 3\slj{1}{1}{0} and 
2\slj{3}{2}{2} charmonium states at the positions of $Y(3943)$ and 
$Z(3931)$, in order to sharpen our expectations for the properties of 
other states above charm threshold. We also extend our previous 
calculations to include additional states at higher masses. We 
highlight the search for narrow structures in charmed 
meson--anticharmed meson pairs, and remark on analogue states to be 
expected in the $b\bar{b}$ spectrum.

\section{Experimental Status \label{sec:status}}
Let us briefly summarize what experiment has taught us about the new 
levels associated with charmonium, and how the experimental facts 
square with theoretical expectations.

\subsection{Properties of $\eta_{c}^{\prime}$}%
\label{subsec:etacp}
The hyperfine partner of $\psi(2\mathrm{S})$ now seems convincingly
established, thanks to observations by the
Belle~\cite{Choi:2002na,Abe:2004ww}, CLEO~\cite{Asner:2003wv}, and
BaBar~\cite{Aubert:2003pt} experiments.  The world-average values of
the mass and width are $M(\eta_{c}^{\prime}) = 3638 \pm 5\mev$ and
$\Gamma(\eta_{c}^{\prime}) = 14 \pm 7\mev$~\cite{PDG05}.

 The proximity of $\psi^{\prime}$ and $\eta_{c}^{\prime}$ to charm
 threshold means that communication with open-charm channels will alter
 the properties of these two narrow levels.  As we elaborated in
 Ref.~\cite{Eichten:2004uh}, the basic coupled-channel interaction is
 spin-independent, but the unequal masses and quantum numbers of the
 2\slj{3}{1}{1} and 2\slj{1}{1}{0} potential-model eigenstates mean
 that their interaction with different charmed-meson--anticharmed-meson
 channels induces spin-dependent forces that affect the charmonium
 states.  These spin-dependent forces give rise to S-D mixing that
 contributes to the $\psi(3770)$ electronic width, for example, and are
 a source of additional spin splitting.

In a single-channel nonrelativistic potential picture, the 2S hyperfine
splitting is given by $M(\psi^{\prime}) - M(\eta_{c}^{\prime}) =
32\pi\alpha_{s}|\Psi(0)|^{2}/9m_{c}^{2}$.  Normalizing to the observed
1S hyperfine splitting, $M(\jpsi) - M(\eta_{c}) = 117\mev$, we would
find $M(\psi^{\prime}) - M(\eta_{c}^{\prime}) = 67\mev$, to be compared
with the observed $48 \pm 5\mev$ separation.  The 2S induced shifts
calculated in Ref.~\cite{Eichten:2004uh} draw $\psi^{\prime}$ and
$\eta_{c}^{\prime}$ closer by $20.9\mev$, substantially improving the
agreement between theory and experiment.  It is tempting to conclude
that the $\psi^{\prime}$-$\eta_{c}^{\prime}$ splitting reflects the
influence of virtual decay channels.

\subsection{CLEO's $h_{c}$ candidate}\label{subsec:hsubc}
Like the $\eta_{c}^{\prime}$, the spin-singlet $L=1$ level of
charmonium has been an object of desire for three decades.  The barrier
to observation has been the absence of allowed E1 or M1 transitions
from the $J^{PC} = 1^{--}$ states readily formed in $e^{+}e^{-}$
annihilations.  The 1\slj{1}{2}{1} state holds a special interest,
beyond the tidiness that would come from completing the spectrum of
narrow states.  In the nonrelativistic potential-model descriptions
that have provided such a reliable guide to quarkonium spectroscopy,
the central potential is generally taken to consist of a Coulomb piece
with vector Lorentz structure (arising from one-gluon exchange) plus a
confining term with scalar Lorentz structure.  For such a potential,
the hyperfine splitting between spin-singlet and spin-triplet orbital
excitations is very small.  A significant deviation from the
expectation that $M(1\slj{1}{2}{1}) \approx \langle M(1\slj{3}{2}{J})
\rangle$ could offer evidence for an unexpected Lorentz structure, or 
for the importance of effects not included in the nonrelativistic
potential-model description.

An earlier observation of $\bar{p}p \to h_{c} \to \pi^{0}\eta_{c}$ near
the 1\slj{3}{2}{J} centroid in ~Fermilab Experiment
760~\cite{Armstrong:1992ae} is not sustained by two additional data
runs with $2\times$ and $3\times$ the luminosity in the successor
experiment E835~\cite{e835,Joffe:2005ce}.  E835 does report a narrow
($\Gamma < 1\mev$) event excess in the $\bar{p}p \to h_{c} \to
\gamma\eta_{c}$ channel at a mass $M(h_{c}) = 3525.8 \pm 0.2 \pm
0.2\mev$~\cite{e835}. The signal bin contains 13 signal events on a 
3-event background, and the signal strength is $\Gamma(h_{c} \to 
\bar{p}p)\mathcal{B}(h_{c} \to \gamma\eta_{c}) \approx 11\ev$.

The CLEO Collaboration has examined 3.08 million $\psi(2\mathrm{S})$
decays, in search of examples of the isospin-violating decay to
$\pi^{0}h_{c}$, followed by $\pi^{0} \to \gamma\gamma$ and $h_{c} \to
\gamma \eta_{c}$~\cite{Rosner:2005ry,Rubin:2005px}.  They find $150 \pm 40$ signal counts for the
inclusive channel, in which $\eta_{c}$ is not reconstructed, and $17.5
\pm 4.5$ signal counts spread over seven hadronic decay channels of
$\eta_{c}$, for a combined significance exceeding $5\sigma$.  In the
inclusive sample, the angular distribution of photons in the $h_{c} \to
\gamma\eta_{c}$ cascade follows the $1 + \cos^{2}\theta$ pattern
characteristic of E1 emission from a spin-1 state.  The CLEO
experimenters infer the 1\slj{1}{2}{1} mass $M(h_{c}) = 3524.4 \pm 0.6
\pm 0.4\mev$, with a product branching fraction
$\mathcal{B}(\psi(2\mathrm{S}) \to \pi^{0}h_{c}) \mathcal{B}(h_{c} \to
\gamma \eta_{c}) = (4.0 \pm 0.8 \pm 0.7) \times 10^{-4}$.

The 1\slj{1}{2}{1} level lies $1.0 \pm 0.6 \pm 0.4\mev$ below the 
spin-triplet centroid, $\langle M(1\slj{3}{2}{J}) \rangle = 3525.36 
\pm 0.06\mev$~\cite{PDG05}, consistent with theoretical 
expectations.  

\subsection{Rare decays of $\psi(3770)$}%
\label{subsec:psipp}
The color-multipole
expansion~\cite{Gottfried:1978gp,Voloshin:1979hc,Yan:1980uh} yields
symmetry relations for hadronic transition rates between charmonium
states, but does not enable calculations of rates from first
principles.  By the Wigner-Eckart theorem for E1-E1 transitions, all
the $1\slj{3}{3}{J} \to \pi\pi\jpsi$ rates should be equal (for
degenerate \slj{3}{3}{J} states), so a reliable determination of the
rate $\Gamma(\psi(3770) \to \pi\pi\jpsi)$ is important for anticipating
the properties of $\psi_{2}$ and $\psi_{3}$.  In our survey of
$B$-meson gateways to  missing charmonium
states~\cite{Eichten:2002qv}, we adopted the value
$\Gamma(1\slj{3}{3}{1} \to \pi\pi\jpsi) \approx 45\kev$.

The CLEO Collaboration has just reported the first high-significance
($11.6\sigma$) observation of the decay $\psi(3770) \to
\pi^{+}\pi^{-}\jpsi$~\cite{Adam:2005mr}.  Their measured branching
fraction, $\mathcal{B}(\psi(3770) \to \pi^{+}\pi^{-}\jpsi) = (1.89\pm
0.20 \pm 0.20)\times 10^{-3}$, corresponds to a partial width
$\Gamma(1\slj{3}{3}{1} \to \pi^{+}\pi^{-}\jpsi) \approx (45 \pm
9)\kev$.  [The CLEO rate is consistent with, but much more precise
than, the Beijing Spectrometer (BES) observation~\cite{Bai:2003hv},
$\mathcal{B}(\psi(3770) \to \pi^{+}\pi^{-}\jpsi) = (3.4 \pm
1.4 \pm 0.9)\times 10^{-3}$.]
A less significant ($3.4\sigma$) observation of the decay
$\psi(3770) \to \pi^{0}\pi^{0}\jpsi$ is consistent with the expectation
that the $\pi^{0}\pi^{0}\jpsi$ rate should be half the
$\pi^{+}\pi^{-}\jpsi$ rate. Accordingly, we shall take 
$\Gamma(1\slj{3}{3}{J} \to \pi\pi\jpsi) = (68 \pm 15)\kev$ as an 
improved estimate of the dipion decay rates of the $\psi_{2}$ and 
$\psi_{3}$. The increase from our earlier estimate does not alter the 
expectation that the most prominent decay of the 1\slj{3}{3}{2} state 
should be $\psi_{2} \to \gamma \chi_{c1}$.

CLEO has also searched for rare radiative decays of $\psi(3770)$, 
reporting the preliminary values $\Gamma(\psi(3770) \to \gamma 
\chi_{c1}) = (78 \pm 19)\kev$ and $\Gamma(\psi(3770) \to \gamma 
\chi_{c2}) < 20\kev$~\cite{millerlis}. These observations are in line 
with expectations ($\Gamma(1\slj{3}{3}{1} \to \gamma\chi_{c1}) 
\approx 59\kev$ and $\Gamma(1\slj{3}{3}{1} \to \gamma\chi_{c2}) \approx   
4\kev$~\cite{Eichten:2004uh}), and give us no reason to question the 
estimated E1 transition rates for the other 1D levels (see 
\S\ref{subsec:radtran}).

\subsection{Properties of $X(3782)$}%
\label{subsec:xprop}
As the best studied of the new $c\bar{c}$-associated states, $X(3872)$
has been subjected to a broad range of diagnostic tests.  Upon
discovery, $X(3872)$ seemed a likely---though somewhat
heavy---candidate for $\psi_{2}$ (or perhaps $\psi_{3}$), but the
expected radiative transitions to $\chi_{c}$ states have never been
seen.  For a state that lies approximately $50\mev$ above charm
threshold, the narrow width ($\Gamma(X(3872) <
2.3\mev$)~\cite{Choi:2003ue} and the absence of a $D\bar{D}$
signal~\cite{Abe:2003zv} suggest unnatural parity $P = (-1)^{J+1}$.
The BaBar Collaboration has set limits on the existence of a charged 
partner~\cite{Aubert:2004kq}.
In 1.96-TeV $\bar{p}p$ collisions, the production characteristics of 
$X(3872)$ are similar to those of $\psi(2\mathrm{S})$~\cite{Abazov:2004kp}. 
CDF has determined the fraction of $X(3872)$ arising from $B$ decays 
as $16.9 \pm 4.9 \pm 2.0\%$~\cite{cdfxlife}.

The $\pi\pi$ mass spectrum favors high dipion masses~\cite{cdfxmass},
suggesting a $\jpsi\,\rho$ decay that would be incompatible with the
identification of $X(3872) \to \pi^{+}\pi^{-}\,\jpsi$ as the strong
decay of a pure isoscalar state.  See, however, the interesting
discussion of alternative interpretations in Ref.~\cite{BauerPanic}.
Suzuki has suggested~\cite{Suzuki:2005ha} that the $\rho$-like behavior
of the $\pi^+\pi^-$ mass spectrum in these decays may result from a
dominant decay to $\omega \jpsi$ slightly off mass shell and the small
(isospin breaking) $ \omega$-$\rho$ mixing.  In this case the $X(3872)$
could be an isoscalar state as expected in a charmonium interpretation.

Observing---or limiting---the $\pi^{0}\pi^{0}\,\jpsi$ decay remains an
important goal~\cite{Barnes:2003vb}.  An observed $\jpsi\,\pi^{+}\pi^{-}\pi^{0}$ decay
suggests an appreciable transition rate to
$\jpsi\,\omega$~\cite{Abe:2005ix}.  Belle's $4.4$-$\sigma$ observation
of the decay $X(3872) \to \jpsi\,\gamma$~\cite{Abe:2005ix} determines
$C=+$, opposite to the charge-conjugation of the leading charmonium
candidates.  Finally, an analysis of angular
distributions~\cite{Rosner:2004ac} supports the assignment $J^{PC} =
1^{++}$~\cite{Abe:2005iy}, but the mass of $X(3872)$ is too low to be
gracefully identified with the 2\slj{3}{2}{1} charmonium state,
especially if $Z(3931)$ (cf. \S~\ref{subsubsec:z3930}) 
is to be identified as the 2\slj{3}{2}{2} level. The $\chi_{c1}^{\prime}$ 
also lacks an isospin-allowed $\pi\pi\jpsi$ decay mode, and acquires a 
substantial $D\bar{D}^{*}$ width if $M(\chi_{c1}^{\prime}) > 
M(D^{0}) + M(D^{*0})$. We shall examine this hypothesis further in 
\S\ref{subsubsec:x23p1}.

If $X(3872)$ is not a charmonium level, what might it
be~\cite{Swanson:2005tq}?  Three interpretations take the
near-coincidence of the new state's mass and the $D^{0}\bar{D}^{*0}$ to
be a decisive clue: an $s$-wave cusp at $D^{0}\bar{D}^{*0}$
threshold~\cite{Bugg:2004rk}, a $D^{0}$ -- $\bar{D}^{*0}$ ``molecule''
bound by pion exchange~\cite{Tornqvist:2004qy,swanghp,Voloshin:2003nt},
and a diquark--antidiquark ``tetraquark'' state
$[cq][\bar{c}\bar{q}]$~\cite{Maiani:2004vq,Maiani:2005pe}.  According
to a different four-quark interpretation, $X(3872)$ is a 
$c\bar{c}q\bar{q}$ state organized by the chromomagnetic 
(color-hyperfine) interaction into a 
$(c\bar{c})_{\mathbf{8}}$-$(q\bar{q})_{\mathbf{8}}$ pair~\cite{Hogaasen:2005jv}.

What distinctive predictions might allow us to put these
interpretations to the test?  On the threshold enhancement
interpretation, we should expect bumps at many thresholds, but no
radial or orbital excitations.  If pion exchange is decisive, then
there should be no analogue molecule at $D_{s}\bar{D}_{s}^{*}$
threshold.  The tetraquark interpretation suggests that $X(3872)$
should be split into two levels, because $[cu][\bar{c}\bar{u}]$ and
$[cd][\bar{c}\bar{d}]$ would be displaced by about $7\mev$.  A first
experimental test by the BaBar Collaboration is
inconclusive~\cite{Aubert:2005zh}.  In $61.2 \pm 15.3$ events that fit
the hypothesis $B^{-} \to K^{-}X(3872)$, they determine a mass of
$3871.3 \pm 0.6 \pm 0.1\mev$, whereas $8.3 \pm 4.5$ $B^{0} \to
K^{0}X(3872)$ events yield $3868.6 \pm 1.2 \pm 0.2\mev$.  The mass
difference, $2.7 \pm 1.3 \pm 0.2\mev$, doesn't yet distinguish between
one $X$ and two.  If diquarks are useful dynamical objects, there
should be a sequence of excited states as well. On the 
color-hyperfine $c\bar{c}q\bar{q}$ interpretation, the only analogous 
level should be a $(b\bar{b})_{\mathbf{8}}$-$(q\bar{q})_{\mathbf{8}}$ 
state near $B\bar{B}^{*}$ threshold.

 What if the $D^{0}\bar{D}^{*0}$ threshold is not the decisive element?
 Hybrid $c\bar{c}$-gluon states might appear in the charmonium
 spectrum~\cite{Close:2003mb}.  The mass and $1^{++}$ quantum numbers
 of $X(3872)$ do not match lattice-QCD
 expectations~\cite{Juge:1999ie,Liao:2002rj,Juge:2002br}.  The valence
 gluon in the hybrid charmonium wave function leads to speculation that
 the $\eta\jpsi$ decay mode might be quite prominent.  The BaBar
 experiment~\cite{Aubert:2004fc} has found no sign of $X(3872) \to
 \eta\jpsi$.  More generally, it is plausible that the gluonic degrees
 of freedom should manifest themselves as ``vibrational states'' in the
 charmonium spectrum~\cite{Giles:1977mp,Buchmuller:1979gy}.  While
 theoretical estimates of the spectrum of string-vibration modes do not
 reproduce the characteristics of $X(3872)$, we should be alert to the
 appearance of such states elsewhere.

\subsection{Evidence for states beyond $X(3782)$}\label{subsec:higherstates}
The other new states associated with charmonium require confirmation 
and elaboration. For the moment, experiments have identified four 
distinct candidates.

\subsubsection{$Y(3940)$}\label{subsubsec:y3940}
Exploring the $\jpsi\,\omega$ mass spectrum in decays $B \to K 
\jpsi\,\omega$, the Belle Collaboration observed an enhancement of $58 
\pm 11$ events near threshold~\cite{Abe:2004zs}. Treated as an 
$s$-wave resonance, this state, named $Y(3940)$, has a mass $M(Y) = 
3943 \pm 11 \pm 13\mev$ and a total width $\Gamma = 87 \pm 22 \pm 
26\mev$.  The mass of this object 
lies well above $DD^{*}$ threshold, so if $Y(3940)$ is a charmonium 
state, it would be expected to decay dominantly into $D\bar{D}$ or 
$D\bar{D}^{*}$. A small $\jpsi\,\omega$ decay rate would 
complicate the interpretation of the product branching fraction,
$\mathcal{B}(B \to KY(3940))\mathcal{B}(Y(3940) \to \jpsi\,\omega) = 
(7.1 \pm 1.3 \pm 3.1) \times 10^{-5}$. For comparison, the branching 
fractions for the decays $B^{+} \to K^{+}(c\bar{c})$, with 
$(c\bar{c}) = \eta_{c}, \jpsi, \chi_{c0}, \chi_{c1}, \psi^{\prime}$, 
are between $6\hbox{ -- }10\times 10^{-4}$~\cite{PDG05}. This suggests a branching 
ratio of about 10\% for $Y(3940) \to \omega\jpsi$ if $Y(3940)$ is a 
conventional charmonium state. In contrast, a 
$c\bar{c}$-gluon hybrid state might decay preferentially into 
$\jpsi$ or $\psi^{\prime}$ plus light 
hadrons~\cite{Close:1994zq,Close:2005iz}.   To settle the identity 
of $Y(3940)$ will require confirming that it is a single state and 
distinguishing it from the $X(3940)$, to which we now turn.

\subsubsection{$X(3940)$}\label{subsubsec:x3940}
The Belle~\cite{Abe:2002rb,Abe:2004ww} and BaBar 
Collaborations~\cite{Aubert:2005tj}  have observed copious double 
charmonium production in $e^{+}e^{-}$ annihilations, with strong 
signals for $\jpsi\eta_{c}$, $\jpsi \chi_{c0}$, and $\jpsi 
\eta_{c}^{\prime}$. Searching in the region above $D\bar{D}$ 
threshold, Belle has observed a new peak at $3943 \pm 6 \pm 6\mev$, 
with a width $\Gamma < 52\mev$ at 90\% C.L.~\cite{Abe:2005hd}.

The new structure, given the temporary name $X(3940)$, is seen to 
decay into $D\bar{D}^{*}$, with a branching fraction 
$\mathcal{B}(X(3940) \to D\bar{D}^{*}) = 0.96^{+0.45}_{-0.32} \pm 0.22$. 
The absence of any signal in the $D\bar{D}$ channel (for which the 
branching fraction is $< 0.41$ at 90\% C.L.) suggests that $X(3940)$ 
is an unnatural parity state. Its properties roughly match those 
expected for $\eta_{c}(3\mathrm{S})$, and we shall examine that 
hypothesis further in \S\ref{subsec:etacpp}.

\subsubsection{$Z(3930)$}\label{subsubsec:z3930}
The Belle Collaboration has examined the $D\bar{D}$ invariant-mass
distribution in $\gamma\gamma$-fusion events in which the transverse
momentum of the $D\bar{D}$ pair is small.  A narrow peak of $64 \pm 18$
events appears at $M(D\bar{D}) = 3939 \pm 5 \pm
2\mev$~\cite{Uehara:2005qd}.  The state, provisionally named $Z(3930)$,
has a total width $\Gamma(Z(3930)) = 29 \pm 10 \pm 2\mev$, and
$\Gamma(Z(3930) \to \gamma\gamma)\mathcal{B}(Z(3930) \to D\bar{D}) =
0.18 \pm 0.05 \pm 0.03\kev$, assuming $J=2$.  The (helicity) angular
distribution strongly prefers $\sin^{4}\theta^{*}$, which corresponds
to a spin-2 resonance, over the flat angular distribution
characteristic of spin-0.  Belle accordingly proposes to identify
$Z(3930)$ as the 2\slj{3}{2}{2} $\chi_{c2}^{\prime}$.  The measured
properties are in reasonable accord with those anticipated in
Ref.~\cite{Eichten:2004uh}; we probe the 2\slj{3}{2}{2} assignment
further in \S\ref{subsubsec:3p2}.

\subsubsection{$Y(4260)$}\label{subsubsec:y4260}
In a study of initial-state radiation events, $e^+e^- \to (\gamma)
\pi^{+}\pi^{-}\jpsi$, the BaBar Collaboration has observed an
accumulation of events near $4260\mev$ in the $\pi^{+}\pi^{-}\jpsi$
invariant mass distribution~\cite{Aubert:2005rm}.  The excess of $125
\pm 23$ events can be characterized as a single $J^{PC} = 1^{--}$
resonance with mass $4259 \pm 8 ^{+2}_{-6}\mev$ and
total width $\Gamma(Y(4260)) = 88 \pm 23 ^{+6}_{-4}\mev$, but a more
complicated structure has not been ruled out.  It is perhaps 
significant that $Y(4260)$ lies near the 
$D_{s}^{*+}\bar{D}_{s}^{*-}$ threshold at $4224\mev$.
BaBar also reports a $3.1\sigma$ signal for $Y(4260)$ in exclusive
$B^{\pm} \to K^{\pm} \pi^{+}\pi^{-}\jpsi$ decays~\cite{Aubert:2005zh}.
An excess of $128 \pm 42$ signal events corresponds to a product 
branching fraction $\mathcal{B}(B^{-} \to K^{-} Y(4260)) 
\mathcal{B}(Y(4260) \to \pi^{+}\pi^{-}\jpsi) = (2.0 \pm 0.7 \pm 0.2) 
\times 10^{-5}$.

The new state has variously been interpreted as $\psi(4\mathrm{S})$, a
$c\bar{c}g$ hybrid state~\cite{Zhu:2005hp,Kou:2005gt,Close:2005iz}, and the radial
excitation of a diquark-antidiquark state identified with
$X(3872)$~\cite{Maiani:2005pe}.  Curiously, the $\pi^{+}\pi^{-}\jpsi$
enhancement occurs in the neighborhood of a local minimum in the cross
section for electron-positron annihilation into hadrons. We shall have 
more to say about possible interpretations in \S\ref{subsec:Babar}.

\section{Strong Dynamics near Threshold}
Near the threshold for open heavy-flavor pair production, light-quark 
pairs induce significant nonperturbative contributions to the masses, 
wave functions, and decay properties of physical $Q\bar{Q}$ states. 
Lattice QCD calculations, once extended into the flavor-threshold 
region, should provide firm theoretical predictions. At present, only 
a phenomenological approach can offer a detailed description of these 
effects.

The effects of light-quark pairs can be described by coupling the
potential model $Q\bar Q$ states to nearby physical multibody states.
In this threshold picture, the strong interactions are broken into
sectors defined by the number of valence quarks.  We decompose the full
Hamiltonian ${\mathcal H}$ as
\begin{equation} 
    \mathcal{H} =  \sum_n \mathcal{H}_n +  \mathcal{H}_I 
\label{eq:calH}
\end{equation} 
where $\mathcal{H}_n$ is the Hamiltonian that governs the
$(c\bar{c}) + n$-light-quark sector and $\mathcal{H}_I$ is the interaction that
couples the various sectors.

The dynamics of the $Q\bar{Q}$ states (with no valence light quarks,
$q$) is described by the interaction ${\mathcal H}_0$.  In principle,
excited $Q\bar Q$ states with valence or vibrational gluonic degrees of
freedom would also be contained in the spectrum of $\mathcal{H}_0$.  In
practice, a simple nonrelativistic potential model is used to determine
the properties of the bound states in this sector.

The two-meson sector $Q\bar q + q\bar Q$ is described by the
Hamiltonian $\mathcal{H}_2$.  As a first approximation, we shall
assume  $\mathcal{H}_2$ to be represented by the low-lying spectrum of
two free heavy-light mesons.  The physical situation is more complex.
At large separation between the two mesons the interactions are
dominated by $t$-channel pion exchanges, when they are allowed.  For
states very near threshold such as $X(3872)$, such pion exchange in
attractive channels might have significant effects on properties of the
physical states \cite{Braaten:2003he}.  At somewhat shorter distances,
more complicated interactions exist and new bound states might arise,
e.g. molecular states
\cite{Voloshin:1976ap,DeRujula:1976qd,DeRujula:1977fs,Tornqvist:2004qy,swanghp,Voloshin:2003nt}.
  Furthermore, at small $Q\bar Q$
separation the $Q\bar Q + q\bar q$ sector may contribute significantly
to ${\mathcal H}_2$.  In particular, as Swanson has
emphasized~\cite{Swanson:2003tb}, the $\rho\jpsi$ and $\omega\jpsi$
thresholds lie in the $D\bar{D}$ threshold region.  The main landmarks
throughout the region we consider are shown in
Table~\ref{table:thresholds}.
 \begin{table}
 \caption{Thresholds for decay into open charm and nearby 
 hidden-charm thresholds.\label{table:thresholds}}
 \begin{ruledtabular}
     \begin{tabular}{c d}
 Channel & \multicolumn{1}{c}{Threshold Energy (MeV)} \\
 \hline
 & \\[-6pt]
 \multicolumn{1}{c}{$D^{0}\bar{D}^{0}$}  &  
 3729.4 \\
 \multicolumn{1}{c}{$D^{+}D^{-}$} & 
 3738.8  \\
 \multicolumn{1}{c}{$D^{0}\bar{D}^{*0} \hbox{ or } D^{*0}\bar{D}^{0}$} & 
   3871.2  \\
   \multicolumn{1}{c}{$\rho^{0}\jpsi$} & 
   3872.7  \\
 \multicolumn{1}{c}{$D^{\pm}D^{*\mp}$} & 
 3879.5  \\
 \multicolumn{1}{c}{$\omega^{0}\jpsi$} & 
 3879.6  \\
 \multicolumn{1}{c}{$D_{s}^{+}D_{s}^{-}$} &   3936.2 \\
 \multicolumn{1}{c}{$D^{*0}\bar{D}^{*0}$} &  4013.6 \\
 \multicolumn{1}{c}{$D^{*+}D^{*-}$} &  4020.2 \\
 \multicolumn{1}{c}{$\eta^{\prime}\jpsi$} & 
 4054.7  \\
 \multicolumn{1}{c}{$f^{0}\jpsi$} & 
 \approx 4077  \\
 \multicolumn{1}{c}{$D_{s}^{+}\bar{D}_{s}^{*-} \hbox{ or }
 D_{s}^{*+}\bar{D}_{s}^{-}$} &   4080.0  \\
 \multicolumn{1}{c}{$a^{0}\jpsi$} & 
 4081.6 \\
 \multicolumn{1}{c}{$\varphi^{0}\jpsi$} & 
 4116.4  \\
 \multicolumn{1}{c}{$D_{s}^{*+}D_{s}^{*-}$} &   4223.8 \\[3pt]
 \multicolumn{1}{c}{$\Lambda_{c}\bar{\Lambda}_{c}$} &   4569.8 \\
      \end{tabular}
 \end{ruledtabular}
 \end{table}

Because current mastery of nonperturbative quantum chromodynamics does
not suffice to derive a realistic description of the interactions that
link the $Q\bar Q$ and $Q\bar q + q\bar Q$ sectors, we must resort to a
phenomenological \textit{Ansatz} for $\mathcal{H}_I$.  Following our
earlier work~\cite{Eichten:2004uh}, we consider here the Cornell
coupled-channel (\ccc) model
\cite{Eichten:1975ag,Eichten:1978tg,Eichten:1979ms} for the creation of
$q\bar{q}$ pairs.  A brief discussion of various other models for
${\mathcal H}_I$ is contained in the Quarkonium Working Group's CERN
Yellow Report~\cite{Brambilla:2004wf}.  The open-charm threshold region
occupies our interest in this study.  However, analogous effects are
present in the $b\bar b$ states near $B\bar B$ threshold and $c\bar b$
states near $DB$ threshold.  A detailed comparison of different
heavy-quark systems could provide valuable insight into the correct
form for the coupling to light-quark pairs.

The \ccc\ formalism
generalizes the $c\bar{c}$ model 
without introducing new parameters, writing the interaction 
Hamiltonian in second-quantized form as
\begin{equation}
    \mathcal{H}_{I} = \cfrac{3}{8} \sum_{a=1}^{8} 
    \int:\rho_{a}(\mathbf{r}) V(\mathbf{r} - 
    \mathbf{r}^{\prime})\rho_{a}(\mathbf{r}^{\prime}): 
    d^{3}{r}\,d^{3}{r}^{\prime}\; ,
    \label{eq:CCCMH}
\end{equation}
where $V$ is the charmonium potential and $\rho_{a}(\mathbf{r}) =
\cfrac{1}{2}\psi^{\dagger}(\mathbf{r})\lambda_{a}\psi(\mathbf{r})$ is
the color current density, with $\psi$ the quark field operator and
$\lambda_{a}$ the octet of SU(3) matrices.  To generate the relevant
interactions, $\psi$ is expanded in creation and annihilation operators
(for charm, up, down, and strange quarks), but transitions from two
mesons to three mesons and all transitions that violate the Zweig rule
are omitted.  It is a good approximation to neglect all effects of the
Coulomb piece of the potential in (\ref{eq:CCCMH}).  This simple model
for the coupling of charmonium to charmed-meson decay channels gives a
qualitative understanding of the structures observed above threshold
while preserving the successes of the single-channel $c\bar{c}$
analysis below threshold~\cite{Eichten:1978tg,Eichten:1979ms}.

\subsection{Mass Shifts \label{subsec:masshift}}
In the presence of coupling to two-light-quark decay channels, the mass
$\omega$ of the quarkonium state $\Psi$ is defined by the eigenvalue 
equation
\begin{equation} 
    [\mathcal{H}_0 + \mathcal{H}_2 + \mathcal{H}_I ] \Psi = \omega \Psi .
\label{eq:decayM}
\end{equation}
Above the flavor threshold, $\omega$ is a complex eigenvalue. 
 
The basic coupled-channel interaction $\mathcal{H}_I$ given by
(\ref{eq:CCCMH}) is independent of the heavy quark's spin, but the
hyperfine splittings of $D$ and $D^{*}$, $D_{s}$ and $D_{s}^{*}$,
induce spin-dependent forces that affect the charmonium states.  These
spin-dependent forces give rise to S-D mixing that contributes to the
electronic widths of \slj{3}{3}{1} states and induces additional spin
splitting among the physical states.

The masses that result from the full coupled-channel analysis are shown
in the second column of Table~\ref{tbl:delM}, which revises and extends
\begin{table}[tb]
\caption{Charmonium spectrum, including the influence of open-charm 
channels. All masses are in MeV. The penultimate column holds an estimate 
of the spin splitting due to tensor and spin-orbit forces in a 
single-channel potential model. The last 
column gives the spin splitting induced by communication with 
open-charm states, for an initially unsplit multiplet.}
 \label{tbl:delM}
 \begin{ruledtabular}
 \begin{tabular}{ccccc}
 State & Mass & Centroid & $\begin{array}{c} \text{Splitting} 
 \\ \text{(Potential)} \end{array}$ &  $\begin{array}{c} \text{Splitting} 
 \\ \text{(Induced)} \end{array}$ \\
 \hline
 & & & & \\[-9pt]
 $\begin{array}{c}
     1\slj{1}{1}{0} \\
     1\slj{3}{1}{1} 
 \end{array}$ &
 $\begin{array}{c} 2\,979.9\footnotemark[1]  \\ 3\,096.9\footnotemark[1]  
 \end{array}$ & $3\,067.6\footnotemark[2]$ & $\begin{array}{c} -90.5\footnotemark[5] 
 \\ +30.2\footnotemark[5] \end{array}$
    &  $\begin{array}{c} +2.8  \\ -0.9 \end{array} $\\ & & & & \\[-6pt]
 $\begin{array}{c} 
 1\slj{3}{2}{0} \\
 1\slj{3}{2}{1} \\
 1\slj{1}{2}{1} \\
 1\slj{3}{2}{2}  \end{array}$ &  
 $\begin{array}{c}
 3\,415.3\footnotemark[1]\\
 3\,510.5\footnotemark[1]\\
 3\,524.4\footnotemark[6]\\
 3\,556.2\footnotemark[1] \end{array}$
 & $3\,525.3\footnotemark[3]$ &  $\begin{array}{c} 
 -114.9\footnotemark[5] \\ -11.6\footnotemark[5] 
 \\ +0.6\footnotemark[5] \\ +31.9\footnotemark[5] \end{array}$ & 
 $\begin{array}{c} +5.9 \\ -2.0 \\ +0.5 \\ -0.3
 \end{array}$ \\
 & & & & \\[-6pt]
 $\begin{array}{c}
     2\slj{1}{1}{0} \\
     2\slj{3}{1}{1} 
 \end{array}$ &
 $\begin{array}{c} 3\,638\footnotemark[1] \\ 3\,686.0\footnotemark[1]
 \end{array}$ & $3\,674\footnotemark[2]$ & $\begin{array}{c} -50.1\footnotemark[5] 
 \\ +16.7\footnotemark[5] \end{array}$
    &  $\begin{array}{c} +15.7  \\ -5.2 \end{array} $\\
    & & & & \\[-6pt]
 $\begin{array}{c} 
 1\slj{3}{3}{1} \\
 1\slj{3}{3}{2} \\
 1\slj{1}{3}{2} \\
 1\slj{3}{3}{3}  \end{array}$ & 
 $\begin{array}{c}
 3\,769.9\footnotemark[1]\\
 3\,830.6\\
 3\,838.0\\
 3\,868.3 \end{array}$
 & 
 (3\,815)\footnotemark[4] & 
 $\begin{array}{c} -40 \\ 0 \\ 
 0 \\ +20 \end{array} $ & 
 $\begin{array}{c} -39.9 \\ -2.7\\ +4.2 \\ +19.0
 \end{array}$ \\ & & & & \\[-6pt]
 $\begin{array}{c} 
 2\slj{3}{2}{0} \\
 2\slj{3}{2}{1} \\
 2\slj{1}{2}{1} \\
 2\slj{3}{2}{2}  \end{array}$ & 
 $\begin{array}{c}
 3\,881.4\\
 3\,920.5\\
 3\,919.0\\
 3\,931\footnotemark[7] \end{array}$
 & (3\,922)\footnotemark[4] & 
 $\begin{array}{c} -90 \\ -8 \\ 
 0 \\ +25 \end{array} $ & 
 $\begin{array}{c} +27.9  \\ +6.7 \\ -5.4  \\ -9.6
 \end{array}$ \\
 & & & & \\[-6pt]
 $\begin{array}{c}
     3\slj{1}{1}{0}\\
     3\slj{3}{1}{1} 
 \end{array}$ & 
 $\begin{array}{c} 3\,943\footnotemark[8] \\ 4\,040\footnotemark[1]
 \end{array}$ & $(4\,015)\footnotemark[9]$ & $\begin{array}{c} -66\footnotemark[5]
 \\ +22\footnotemark[5] \end{array}$
    &  $\begin{array}{c} -3.1  \\ +1.0\end{array} $\\
 \end{tabular}
 \end{ruledtabular}
 \footnotetext[1]{Observed mass, from \textit{Review of Particle 
 Physics,} Ref.~\cite{PDG05}.}
 \footnotetext[2]{Input to potential determination.}
 \footnotetext[3]{Observed 1\slj{3}{2}{J} centroid.}
 \footnotetext[4]{Computed centroid.}
 \footnotetext[5]{Required to reproduce observed masses.}
 \footnotetext[6]{Observed mass from CLEO \cite{Rubin:2005px}.}
 \footnotetext[7]{Observed mass from Belle \cite{Uehara:2005qd}.} 
 \footnotetext[8]{Observed mass from Belle \cite{Abe:2005hd}.}
 \footnotetext[9]{Observed 3S centroid.}
 \end{table}
 our previously published results~\cite{Eichten:2004uh}.  The new
 version presented here includes the 3S levels and takes account of
 Belle's evidence~\cite{Uehara:2005qd} for $Z(3930)$, interpreted as a
 2\slj{3}{2}{2} state (cf.  \S\ref{subsubsec:z3930}).  As in our
 earlier analysis, the parameters of the potential-model sector
 governed by ${\cal H}_0$ must be readjusted to fit the physical
 masses, $\omega$, to the observed experimental values.  The centroids
 of the 1D and 2P spin-triplet masses are pegged to the observed masses
 of 1\slj{3}{3}{1} [$\psi(3770)$] and 2\slj{3}{2}{2} [$Z(3930)$],
 respectively.  The assumed spin splittings in the single-channel
 potential model are shown in the penultimate column and the induced
 coupled-channel spin splittings for initially unsplit multiplets are
 presented in the rightmost column of Table~\ref{tbl:delM}.  The shifts
 induced in the low-lying 1S and 1P levels are small.  For all the
 other states, coupled-channel effects are noticeable and interesting.

An important consequence of coupling to the open-charm threshold is that
the $\psi^{\prime}$ receives a downward shift through its communication
with the nearby $D\bar{D}$ channel; the unnatural parity
$\eta_{c}^{\prime}$ does not couple to $D\bar{D}$, and so is not
depressed in the same degree.  This effect is implicitly present in the
early Cornell papers~\cite{Eichten:1978tg, Eichten:1979ms}, but the
shift of spin-singlet states was not calculated there.  The first
explicit mention---and the first calculation---of the unequal effects
on the masses of the 2S hyperfine partners is due to Martin and
Richard~\cite{Martin:1982nw}.  In the framework of the
\ccc\ model, we found~\cite{Eichten:2002qv, Eichten:2004uh} (cf. 
Table~\ref{tbl:delM}) that the
induced shifts draw $\psi^{\prime}$ and
$\eta_{c}^{\prime}$ closer by $20.9\mev$, substantially improving the
agreement between theory and experiment.  This suggests that the
$\psi^{\prime}$-$\eta_{c}^{\prime}$ splitting reflects the influence of
virtual decay channels.  In the case of the 3S system, both the
3\slj{1}{1}{0} $\eta_{c}^{\prime\prime}$ and the 3\slj{3}{1}{1}
$\psi(4040)$ communicate with open decay channels, and the \ccc\ model
leads to a modest 9-MeV \textit{increase} in the interval between them, as seen
in Table \ref{tbl:delM}.

\subsection{Mixing and Properties of Physical States 
\label{subsec:mixphys}}

The physical states are not pure potential-model eigenstates but
include components with two virtual (real, above threshold) open-flavor
meson states.  Separating the physical state $\Psi$ into $Q\bar Q$
($\Psi_0$) and two-charmed-meson components ($\Psi_2$), the resulting
decomposition of $\cal H$ by sector leads to an effective Hamiltonian for the
$Q\bar{Q}$ sector given by
\begin{equation} 
\left[ {\cal H}_0 + {\cal H}_I^{\dagger} \frac{1}{\omega - {\cal H}_2 + i\epsilon}
{\cal H}_I \right] \Psi_0 = \omega \Psi_0 \; .
\label{eq:redH} 
\end{equation}

 Solving Eqn.~(\ref{eq:redH}) in the $Q\bar Q$ sector determines the
 mixing among the potential-model states and the coupling to decay
 channels.  Calculational procedures for the \ccc\ model based on the
 coupling Hamiltonian (\ref{eq:CCCMH}) have been described in detail
 elsewhere~\cite{Eichten:1978tg,Eichten:1979ms,Eichten:2004uh}.  Even
 above threshold for strong decays, the coupled-channel Hamiltonian
 approach (\ref{eq:redH}) allows a definition of the mass and width of
 the resonances in terms of the complex eigenvalues and a decomposition
 of the state into the $c\bar c$ and $D\bar D$ components via the
 associated eigenvector.  Of course, the width and mass so defined will
 differ somewhat from the resonance mass and width inferred from
 the observed enhancement in any particular open-charm--meson-pair
 channel.  

The two-meson contributions to the wave functions of the low-lying
$c\bar c$ states are shown in Table~\ref{tbl:Zf} for the \ccc\ model.
The overall probability for the physical state to be in what we have
called the $\Psi_{0}$ $(c\bar c)$ sector, denoted $\zcc$, decreases as
\begin{table}[tb] 
\caption{Wave function fractions (in percent) in the $\Psi_{0}$
$(c\bar{c})$ and $\Psi_{2}$ ($D^{(*)}\bar{D}^{(*)}$) sectors for 
near-threshold states.}
\label{tbl:Zf} 
\begin{ruledtabular} 
\begin{tabular}{cc|cccc} 
State & $\Psi_{0}$ fraction & \multicolumn{4}{c}{$\Psi_{2}$ fraction} \\ 
& $\zcc$ &  & $D\bar D$ & $D\bar D^* + D^*\bar D$ & $D^*\bar D^*$  \\ 
\hline 
 & & & & & \\[-6pt]
$~1\slj{3}{3}{1}~$ & $~50.03~$ & $~u~$ & $17.76$ & $4.43$ & $2.42$ \\
	     & $~    ~$ & $~d~$ & $15.83$ & $4.18$ & $2.38$ \\
	     & $~    ~$ & $~s~$ & $1.33$ & $0.85$ & $0.79$ \\
\hline
$~1\slj{3}{3}{2}~$ & $~61.10~$ & $~u~$ & $0.00$ & $15.10$ & $3.30$ \\
		       & $~    ~$ & $~d~$ & $0.00$ & $13.94$ & $3.21$ \\
		       & $~    ~$ & $~s~$ & $0.00$ & $2.38$ & $0.96$ \\
 \hline
 $~1\slj{1}{3}{2}~$ & $~63.36~$ & $~u~$ & $0.00$ & $12.33$ & $4.88$ \\
		       & $~    ~$ & $~d~$ & $0.00$ & $11.38$ & $4.74$ \\
		       & $~    ~$ & $~s~$ & $0.00$ & $2.00$ & $1.31$ 
		       \\
\hline
$~1\slj{3}{3}{3}~$ & $~50.06~$ & $~u~$ & $16.51$ & $1.59$ & $7.47$ \\
		       & $~    ~$ & $~d~$ & $13.17$ & $1.54$ & $7.20$ \\
		       & $~    ~$ & $~s~$ & $0.25$ & $0.43$ & $1.77$ \\
		       \hline
$~2\slj{3}{2}{0}~$ & $~50.03~$ & $~u~$ & $16.48$ & $0.00$ & $5.23$ \\
		       & $~    ~$ & $~d~$ & $17.74$ & $0.00$ & $5.04$ \\
		       & $~    ~$ & $~s~$ & $4.41$ & $0.00$ & $1.06$ \\
		       \hline
 $~2\slj{1}{2}{1}~$ & $~50.04~$ & $~u~$ & $0.00$ & $19.40$ & $4.95$ \\
		       & $~    ~$ & $~d~$ & $0.00$ & $18.73$ & $4.68$ \\
		       & $~    ~$ & $~s~$ & $0.00$ & $1.43$ & $0.78$ 
		       \\
		       \hline
 $~2\slj{3}{2}{1}~$ & $~50.06~$ & $~u~$ & $0.00$ & $20.82$ & $3.10$ \\
		       & $~    ~$ & $~d~$ & $0.00$ & $20.41$ & $2.99$ \\
		       & $~    ~$ & $~s~$ & $0.00$ & $2.00$ & $0.63$ \\
		       \hline
 $~2\slj{3}{2}{2}~$ & $~49.95~$ & $~u~$ & $4.47$ & $12.80$ & $7.74$ \\ 
		       & $~    ~$ & $~d~$ & $4.32$ & $11.47$ & $7.17$ \\
		       & $~    ~$ & $~s~$ & $0.24$ & $0.87$ & $0.97$ \\
		       \hline
 $~3\slj{1}{1}{0}~$ & $~50.05~$ & $~u~$ & $0.00$ & $20.78$ & $4.02$ \\
		       & $~    ~$ & $~d~$ & $0.00$ & $19.82$ & $3.78$ \\
		       & $~    ~$ & $~s~$ & $0.00$ & $1.01$ & $0.55$ 
		       \\
		       \hline
$~3\slj{3}{1}{1}~$ & $~50.04~$ & $~u~$ & $1.42$ & $10.12$ & $11.86$ \\
		       & $~    ~$ & $~d~$ & $1.45$ & $10.24$ & $11.24$ \\
		       & $~    ~$ & $~s~$ & $0.98$ & $1.58$ & $1.08$ 
		       \\
		       \hline
$~2\slj{3}{3}{1}~$ & $~50.02~$ & $~u~$ & $3.94$ & $5.34$ & $12.27$ \\
	     & $~    ~$ & $~d~$ & $3.87$ & $5.49$ & $11.89$ \\
	     & $~    ~$ & $~s~$ & $2.69$ & $3.03$ & $1.46$ \\
\end{tabular} 
\end{ruledtabular} 
\end{table} 
charm threshold is approached.  For states above threshold the mixing
coefficients become complex, as shown in Table~\ref{tbl:wf}.
Configuration mixing effects modify radiative transition
rates~\cite{Rosner:2001nm,Rosner:2004wy,Eichten:2004uh}, in addition to their
influence on S-D mixing and spin splittings we have already mentioned.
\begin{table}[tbhp] 
\caption[wf]{Charmonium content of states near $c\bar{c}$ threshold. 
The wave function $\Psi$ reflects mixing induced through 
communication with
open-charm channels. Unmixed potential-model eigenstates are denoted by 
$\Ket{n\sLj{2s+1}{L}{J}}$. The coefficient of the dominant unmixed state is chosen 
real and positive. Physical states are evaluated at 
 masses given in Table~\ref{tbl:delM}.} 
\label{tbl:wf} 
\begin{ruledtabular} 
\begin{tabular}{lcl} 
State & & Principal $\Psi_{0}$ ($c\bar{c}$) Components \\ 
\hline 
 & & \\[-6pt]
  $\Psi(1\slj{3}{3}{1})$ & $=$ & $0.10e^{+0.59i\pi}\Ket{2{\rm S}}+0.01e^{+0.81i\pi}\Ket{3{\rm S}}$\\
			 &     & $ + 0.69\Ket{1{\rm D}}+0.10e^{+0.86i\pi}\Ket{2{\rm D}}$ \\[3pt] 
  $\Psi(1\slj{3}{3}{2})$ & $=$ & $0.77\Ket{1{\rm D}}-0.10\Ket{2{\rm D}}-0.02\Ket{3{\rm D}}$ \\[3pt] 
  $\Psi(1\slj{1}{3}{2})$ & $=$ & $0.79\Ket{1{\rm D}}-0.10\Ket{2{\rm D}}-0.02\Ket{3{\rm D}}$ \\[3pt] 
  $\Psi(1\slj{3}{3}{3})$ & $=$ & $0.70\Ket{1{\rm D}}-0.09\Ket{2{\rm D}}+0.02e^{-0.98i\pi}\Ket{3{\rm D}}$ \\[3pt] 
  $\Psi(2\slj{3}{2}{0})$ & $=$ & $0.11e^{-0.38i\pi}\Ket{1{\rm P}}+0.70\Ket{2{\rm P}}+0.03e^{+0.56i\pi}\Ket{3{\rm P}}$ \\[3pt] 
  $\Psi(2\slj{1}{2}{1})$ & $=$ & $0.18e^{-0.19i\pi}\Ket{1{\rm P}}+0.68\Ket{2{\rm P}}+0.07e^{+0.77i\pi}\Ket{3{\rm P}}$ \\[3pt] 
  $\Psi(2\slj{3}{2}{1})$ & $=$ & $0.18e^{-0.27i\pi}\Ket{1{\rm P}}+0.68\Ket{2{\rm P}}+0.07e^{+0.69i\pi}\Ket{3{\rm P}}$ \\[3pt] 
  $\Psi(2\slj{3}{2}{2})$ & $=$ & $0.18e^{-0.09i\pi}\Ket{1{\rm P}}+0.66\Ket{2{\rm P}}+0.07e^{+0.85i\pi}\Ket{3{\rm P}}$ \\[3pt] 
  $\Psi(3\slj{1}{1}{0})$ & $=$ & $0.02e^{-0.06i\pi}\Ket{1{\rm S}}+0.20e^{-0.18i\pi}\Ket{2{\rm S}}+0.68\Ket{3{\rm S}}$ \\
			 &     & $+0.05e^{+0.67i\pi}\Ket{4{\rm S}}$\\[3pt] 
  $\Psi(3\slj{3}{1}{1})$ & $=$ & $0.02e^{-0.05i\pi}\Ket{1{\rm S}}+0.19e^{-0.30i\pi}\Ket{2{\rm S}}+0.67\Ket{3{\rm S}}$ \\
			 &     & $+0.07e^{+0.54i\pi}\Ket{4{\rm S}}$ \\ 
  & & $+0.04e^{+0.59i\pi}\Ket{1{\rm D}}+0.04e^{+0.59i\pi}\Ket{2{\rm D}}$ \\[3pt] 
  $\Psi(2\slj{3}{3}{1})$ & $=$ & $0.02e^{+0.60i\pi}\Ket{2{\rm S}}+0.03e^{-0.71i\pi}\Ket{3{\rm S}}$ \\
			 &     & $+0.14e^{-0.50i\pi}\Ket{1{\rm D}}+0.69\Ket{2{\rm D}}$ \\[3pt] 
\end{tabular} 
\end{ruledtabular} 
\end{table} 

\subsection{Zweig-Allowed Strong Decays \label{subsec:zweig}}
Once the mass of a resonance is given, the \ccc\ formalism yields
reasonable predictions for the other resonance properties.  We present
in Table \ref{tbl:decay} our estimates of the strong decay rates for
all the $s$-, $p$-, and $d$-wave charmonium levels that populate the
threshold region below $4200$ MeV, together with what is known from
experiment.
\begin{table}
\caption{Open-charm strong decay modes of the charmonium states near
threshold.   The theoretical widths using the \ccc\
model~\cite{Eichten:2004uh} are shown.}
\label{tbl:decay}
\begin{ruledtabular}
\begin{tabular}{lcccc}
State & $n\sLj{2s+1}{L}{J}$ & Mode & \multicolumn{2}{c}{Decay Width (MeV)} \\ 
      &   &   &  Experiment  & Computed \\ 
\hline
 & & & & \\[-6pt]
$\psi(3770)$ & 1\slj{3}{3}{1}
	  &  $D^0\bar{D}^0$        &                   & $11.8$ \\   
	 & &  $D^+D^-$             &                   & $8.3$ \\   
	 & &  total          & $23.6 \pm 2.7$\footnotemark[2] & $20.1$ \\[4pt]
$\psi(3868)$ & 1\slj{3}{3}{3}
	  &  $D\bar{D}$             &                          & $0.82$ \\   
	 & &  total          &                          & $0.82$ \\[4pt]
$\chi_{c0}^{\prime}(3881)$   & 2\slj{3}{2}{0}
	  &  $D\bar{D}$             &                   & $61.5$ \\
	 & &  total          &                   & $61.5$ \\[4pt]
$h_{c1}^{\prime}(3919)$   & 2\slj{1}{2}{1}
	  &  $D\bar{D}^*$         &                   & $59.8$  \\
	 & &  total          &                   & $59.8$  \\[4pt]
$\chi_{c1}^{\prime}(3920)$   & 2\slj{3}{2}{1}
	  &  $D\bar{D}^*$         &                   & $81.0$  \\
	 & &  total          &                   & $81.0$  \\[2pt]
$\chi_{c2}^{\prime}(3931)$   & 2\slj{3}{2}{2}
	  &  $D\bar{D}$             &                   & $21.5$  \\
	 & &  $D\bar{D}^*$         &                   & $7.1$ \\
	 & &  total          & $29 \pm 10{(\rm stat)} \pm 2{(\rm sys)}$\footnotemark[4] & $28.6$ \\[4pt]
$\eta_c^{\prime\prime}(3943)$  & 3\slj{1}{1}{0} 
	  &  $D\bar{D}^*$         &                   & $49.8$ \\
	 & &  total          & $< 52$\footnotemark[5]   & $49.8$  
	 \\[4pt]
$\psi(4040)$  & 3\slj{3}{1}{1} 
	  &  $D\bar{D}$             &                   & $0.1$   \\
	 & &  $D\bar{D}^*$         &                   & $33.$  \\
	 & &  $D_s\bar{D}_s$     &                   & $8.$   \\
	 & &  $D^*\bar{D}^*$     &                   & $33.$  \\
	 & &  total          & $\left\{ \begin{array}{c}
	 52 \pm 10\footnotemark[2]  \\ 88 \pm 5\footnotemark[3] 
     \end{array}\right\}$	 & $74.$  \\[4pt]
$\psi(4159)$   & 2\slj{3}{3}{1}
	  &  $D\bar{D}$             &                   & $3.2$  \\
	 & &  $D\bar{D}^*$         &                   & $6.9$  \\
	 & &  $D^*\bar{D}^*$     &                   & $41.9$ \\
	 & &  $D_s\bar{D}_s$     &                   & $5.6$  \\
	 & &  $D_s\bar{D}_s^*$   &                   & $11.0$ \\
	 & &  total          & $\left\{\begin{array}{c} 78 \pm 20\footnotemark[2]  
	 \\ 107 \pm 8\footnotemark[3] \end{array}\right \}$
	 & $69.2$ \\
\end{tabular}
\end{ruledtabular}
 \footnotetext[1]{Computed from CLEO branching fractions~\cite{Abe:2005hd}.}
 \footnotetext[2]{\textit{Review of Particle 
 Physics}~\cite{PDG05}.}
 \footnotetext[3]{Reanalysis by Seth~\cite{Seth:2004py}.}
 \footnotetext[4]{Belle~\cite{Uehara:2005qd}.}
 \footnotetext[5]{Belle~\cite{Abe:2005hd}.}
\end{table} 
No new experimental information has come to light about $L \ge 3$ 
levels; the predictions given in Ref.~\cite{Eichten:2004uh} remain 
current. 

The 1\slj{3}{3}{1} state $\psi^{\prime\prime}(3770)$, which lies some
$40\mev$ above charm threshold, calibrates the reasonableness of our
calculated widths. As we noted in Ref.~\cite{Eichten:2004uh}, our 
value of $20.1\mev$ is in excellent agreement
with the world average, $\Gamma(\psi(3770)) = 23.6 \pm
2.7\mev$~\cite{PDG05}.

The results presented here differ in two respects from those of
Ref.~\cite{Eichten:2004uh}.  First, the $\chi_{c2}^{\prime}$ and
$\eta_{c}^{\prime\prime}$ masses have been fixed, if only
provisionally, by experiment.  This results in shifts of the masses and
properties of all the 2P and 3S states as shown in
Tables~\ref{tbl:delM}, \ref{tbl:decay}, and \ref{table:E1}.  Second,
the present approach allows a more detailed extraction of the
composition of the charmonium states above threshold.  These results
are shown in Tables~\ref{tbl:Zf} and \ref{tbl:wf}.

Along with the current PDG values for the total widths of higher
$1^{--}$ $c\bar c$ resonances, we show in Table~\ref{tbl:decay} a
reanalysis of the existing experimental data by
Seth~\cite{Seth:2004py}.

The natural-parity 1\slj{3}{3}{3} state can decay into 
$D\bar{D}$, but its $f$-wave decay is suppressed
by the centrifugal barrier factor. 
Thus the 1\slj{3}{3}{3} may be discovered as 
a narrow $D\bar{D}$ resonance up to a mass of about $4000\mev$.

Barnes, Godfrey, and Swanson~\cite{Barnes:2005pb} have recently
reported extensive calculations of decay widths of higher charmonium
states in the framework of the \slj{3}{2}{0} model for quark-pair
production.  [Shortcomings of both the \ccc\ and \slj{3}{2}{0} models
are assessed in Ref.~\cite{Brambilla:2004wf}.  Detailed comparisons
(e.g. Ackleh, Barnes and Swanson~\cite{Ackleh:1996yt}) between various
light quark pair creation models are highly desirable.]  Unlike the
analysis presented here, their calculation {does not resum} the effects
of coupling to decay channels.  The general scale of resulting decay
rates is similar to those displayed in Table~\ref{tbl:decay}.  For
example, Barnes \textit{et al.} determine $\Gamma(3\slj{3}{1}{1}) =
80\mev$ and $\Gamma(2\slj{3}{3}{1}) = 74\mev$.

\subsection{Radiative Transitions \label{subsec:radtran}}
As Tables~\ref{tbl:Zf} and \ref{tbl:wf} show, the physical charmonium
states are not pure potential-model eigenstates.  To compute the E1
radiative transition rates, we must take into account both the standard
$(c\bar{c}) \to (c\bar{c})\gamma$ transitions and the transitions
between (virtual) decay channels in the initial and final states.
Details of the calculational procedure are given in \S IV.B of
Ref.~\cite{Eichten:1979ms}.  There we also illuminated the 
differences between single-channel and coupled-channel expectations 
for the radiative rates.

Our expectations for E1 transition rates
among spin-triplet levels are shown in Table \ref{table:E1}.  Again the
observed properties of $\psi(3770)$ confirm the reasonableness of the
\ccc\ framework.
\begin{table}
\caption{Calculated and observed rates for E1 radiative transitions 
among charmonium levels. 
\ccc\ model rates include the influence of open-charm channels. Photon 
energies, in MeV, in parentheses.
\label{table:E1}}
\begin{ruledtabular}
\begin{tabular}{lccc} 
        & \multicolumn{3}{c}{Partial Width (keV)} \\
\hline
 & & & \\[-6pt]
$1\slj{3}{3}{1}(3770) \to$ & $\chi_{c2}\,\gamma(208)$ & $\chi_{c1}\,\gamma(251)$ &  
$\chi_{c0}\,\gamma(338)$ \\
 model               & $3.9$       & $59$        & $225$ \\ 
 CLEO~\cite{Coan:2005ps} & $<40$       & $75 \pm 18$ &   $<1100$    \\[6pt]
$1\slj{3}{3}{2}(3831) \to$ & $\chi_{c2}\,\gamma(266)$ & $\chi_{c1}\,\gamma(308)$ & \\
 model               & $45$        & $212$       &       \\[6pt]
$1\slj{3}{3}{3}(3868) \to$ & $\chi_{c2}\,\gamma(303)$ & &                           \\
 model               & $286$       &             &       \\[6pt]
$2\slj{3}{2}{0}(3881)\to$ & $\jpsi\,\gamma(704)$ & $\psi^{\prime}\,\gamma(190)$ & 
$1\slj{3}{3}{1}\,\gamma(110)$ \\ 
 model               & $39$        & $17$        & $1.6$ \\[6pt]
{$2\slj{3}{2}{1}(3920) \to$} & $1\slj{3}{3}{1}\,\gamma(147)$ & $1\slj{3}{3}{2}\,\gamma(86)$ & \\ 
 model               & $5.2$       & $2.6$      &       \\[3pt]
 {$2\slj{3}{2}{1}(3920) \to$}     & $\jpsi\,\gamma(737)$    &   $\psi^{\prime}\,\gamma(227)$ & \\ 
 model               & $15$       & $75$       &       \\[6pt]
$2\slj{3}{2}{2}(3931) \to$ & $1\slj{3}{3}{1}\,\gamma(157)$ & $1\slj{3}{3}{2}\,\gamma(95)$ & 
$1\slj{3}{3}{3}\,\gamma(62)$ \\
 model               & $0.4$       & $3.0$       & $3.6$ \\[3pt]
 $2\slj{3}{2}{2}(3931) \to$     & $\jpsi\,\gamma(775)$ & $\psi^{\prime}\,\gamma(272)$ & \\ 
 model               & $14$        & $120$       &       \\[6pt]
$\psi(4040) \to$         & $2\slj{3}{2}{2}\,\gamma(84)$ & $2\slj{3}{2}{1}\,\gamma(132)$ & 
$2\slj{3}{2}{0}\,\gamma(156)$ \\
 model               & $18$        & $5.4$       & $20$  \\[3pt]
 $\psi(4040) \to$ & $1\slj{3}{2}{2}\,\gamma(456)$ & $1\slj{3}{2}{1}\,\gamma(495)$ & 
$1\slj{3}{2}{0}\,\gamma(577)$ \\
 model               & $12$        & $0.4$       & $0.03$ \\[6pt]
\end{tabular}
\end{ruledtabular}
\end{table}
The radiative decay rates for the 2\slj{3}{2}{J} and 3\slj{3}{1}{1}  
levels are calculated at the predicted masses.
All these rates are small compared to the expected open-charm 
decay rates.

\section{Implications for newfound states \label{sec:impnew}}
\subsection{$X(3940)$ as $\eta_c(3\mathrm{S})$ \label{subsec:etacpp}}
\begin{figure}[tb]
    \BoxedEPSF{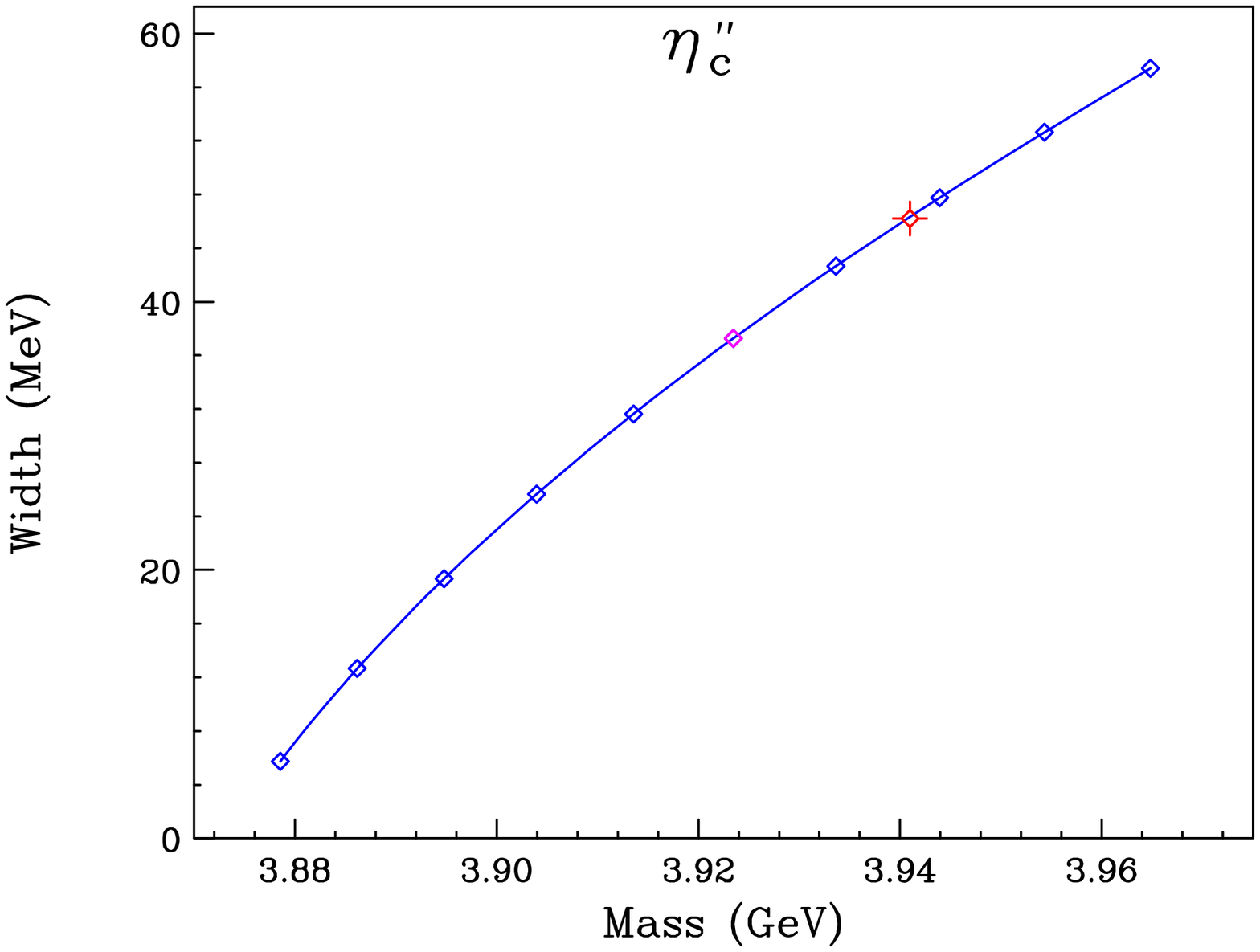 scaled 450}
    \caption{Variation of the $\eta_c^{\prime\prime}$ width with mass
    in the \ccc\ model.  The physical mass of the $X(3940)$ is denoted
    by a crossed dot.\label{fig:uno}}
\end{figure}

The observation by Belle \cite{Abe:2005hd} of a state $X(3940)$
recoiling against the $\jpsi$ system in continuum production at the
$\Upsilon(4\mathrm{S})$ region fits the general behavior expected for the
$\eta_c^{\prime\prime}$ charmonium level.  This state is not seen to decay
into $D\bar D$ but does decay into $D\bar D^*$, suggesting that it has 
unnatural parity.  
Figure~\ref{fig:uno} shows the complex pole position of the 
$\eta_c^{\prime\prime}$ in the \ccc\ model as its bare mass is varied 
in 10-MeV steps. At $3943\mev$, we estimate a $D\bar{D}^{*}$ width 
of $50\mev$, which is quite close to the experimental upper 
bound on the total width, $\Gamma(X(3940)) < 52\mev$. 

\begin{figure}[tb]
    \BoxedEPSF{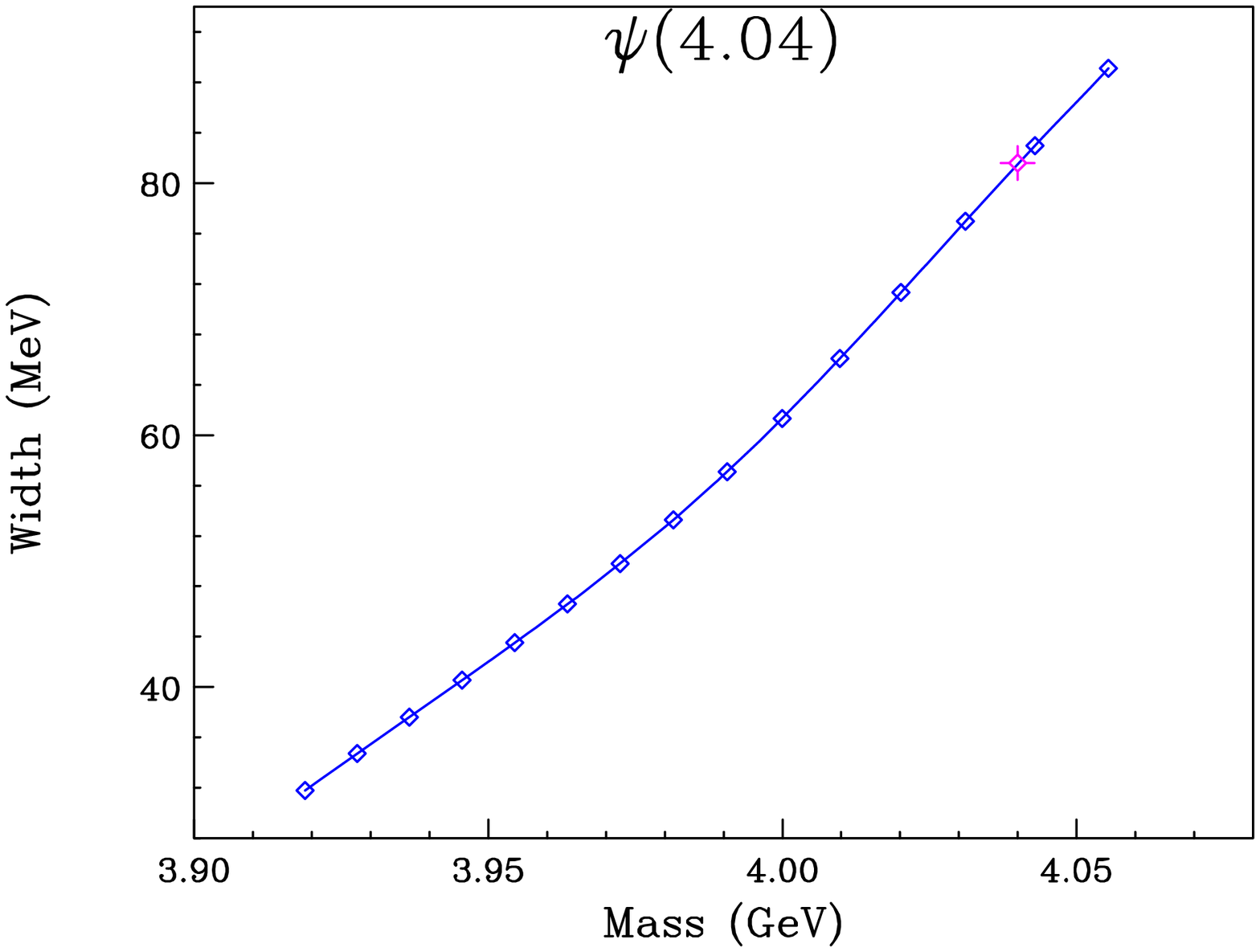 scaled 450}
    \caption{Width of the 3\slj{3}{1}{1} state as a function of its
    mass.  The physical mass of the $\psi(4040)$ is denoted by a
    crossed dot.  \label{fig:due}}
\end{figure}

If $\psi(4040)$ is assigned to the $3\slj{3}{1}{1}$ level, it is
somewhat problematic to identify $X(3940)$ as its hyperfine partner.
We show in Figure~\ref{fig:due} the behavior of the physical mass and
width of the triplet 3S state as the bare mass is varied in 10-MeV
steps.  To reproduce the observed masses, we would require the spin
splitting in the charmonium sector to be $88\mev$, which is
considerably larger than the 2S splitting, and larger than expected in 
naive potential models.

If, for the moment, we take the existence, mass, and unnatural parity
of $X(3940)$ as well established, then two possible sources of
the discrepancy  deserve futher investigation.

First, it is difficult to determine the true pole position for the
$3\slj{3}{1}{1}$ state from measurements of the step in $R_{c\bar{c}}
\equiv \sigma(e^{+}e^{-} \to \mathrm{hadrons})/\sigma(e^{+}e^{-} \to
\mu^{+}\mu^{-})$~\cite{Seth:2004py}.
The resonance decay widths are determined
from fitting measurements of $\Delta R$ in $e^+e^-$ annihilation to a
model for each resonance including radiative corrections.  This whole
procedure is complicated by  its dependence on the resonance shape,
i.e., the expected non-Breit--Wigner nature of the partial widths for
radially excited resonances.  It may be more useful to produce a model
of $\Delta R$ for direct comparison with data.  Greater resolving power
between models is possible if the contribution from each individual
open heavy flavor final state is separately reported.

For the \ccc\ model, the structure of $\Delta R(c\bar c)$ in the
threshold region was studied in the original Cornell group works
\cite{Eichten:1975ag,Eichten:1978tg,Eichten:1979ms} and later extended
to the $\Delta R(b\bar b)$ in the threshold
region~\cite{Eichten:1980ce}.  The structure of $\Delta R(c\bar c)$ and
$\Delta R(b\bar b)$ has also been studied in \slj{3}{2}{0} models of
quark pair creation~\cite{Heikkila:1983wd}.  There are also some
attempts to compare the different
models~\cite{Byers:1989zn,Byers:1994dx}.

Second, thresholds corresponding to an $s$-wave plus a $p$-wave charmed
meson are nearby.  Unlike the two--$s$-wave charm meson channels, which
are $p$-wave decays for the 3S states, these are $s$-wave decays and
thus have stronger threshold effects.  In particular the lowest mass
channel $D(J=0; j_l^P = 1/2^{-}) D(J=0; j_l^P = 1/2^+)$ couples to the
$3^1S_1$ $\eta_c^{\prime\prime}$ state but not the $3^3S_1$
$\psi(4040)$ state.  This will increase the induced spin-splitting 
(see Table~\ref{tbl:delM}, column 5) between the states. Thus the observed 
$\psi^{\prime\prime}\hbox{ - }\eta_{c}^{\prime\prime}$ separation
will correspond to a smaller potential-model splitting (column 4).

\subsection{The 2P states \label{subsec:2P}}
\subsubsection{$Z(3930)$ as $\chi_{c2}^{\prime}$ \label{subsubsec:3p2}}
\begin{figure}[tb]
    \BoxedEPSF{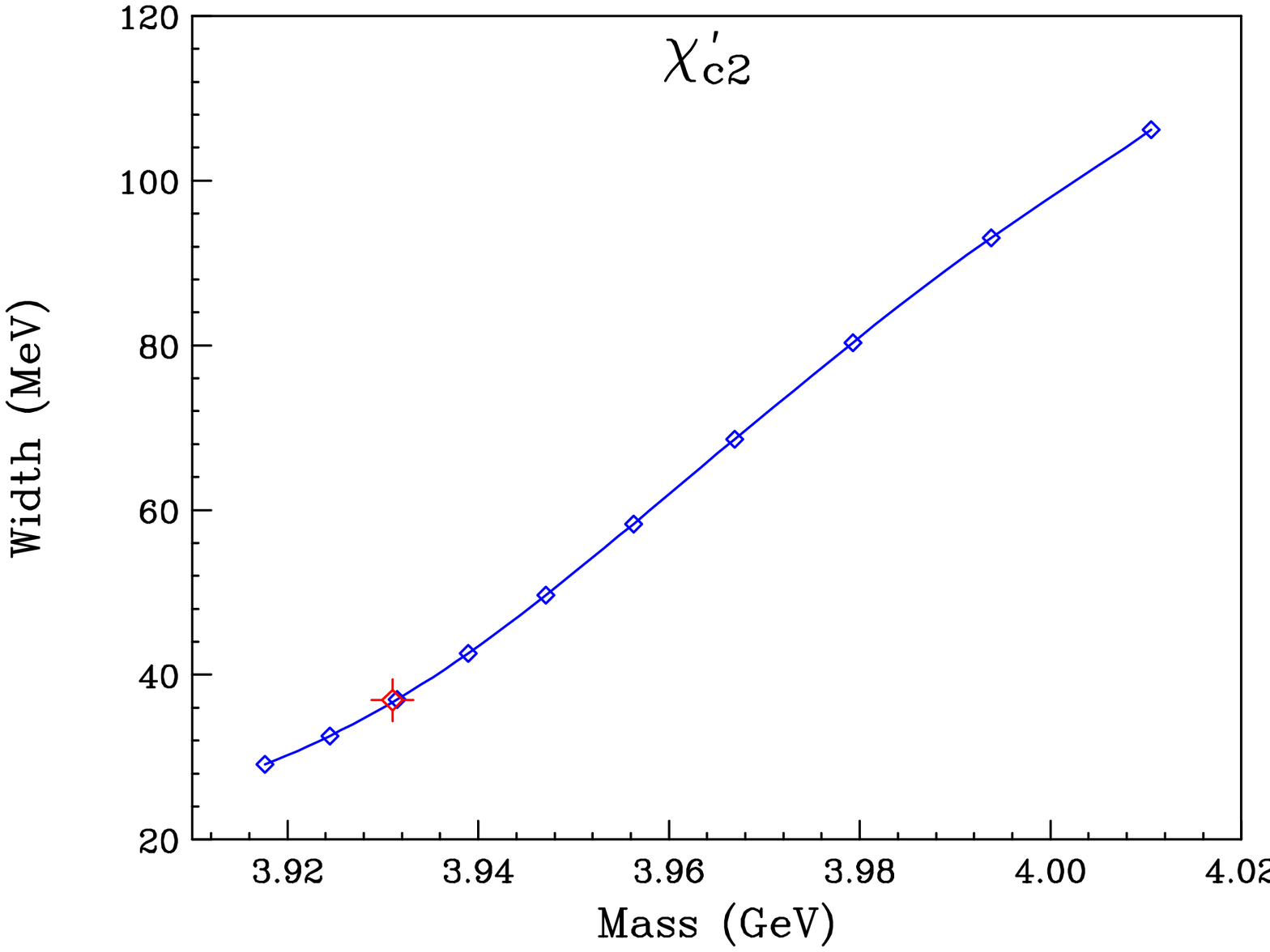 scaled 450}
    \caption{Mass dependence of  $\Gamma(2\slj{3}{2}{2})$. 
    \label{fig:quattro}}
\end{figure}

Belle has also observed the new state $Z(3930)$ in two-photon
production of $D\bar{D}$ pairs~\cite{Uehara:2005qd}.  The 2\slj{3}{2}{2}
and 2\slj{3}{2}{0} levels are natural charmonium candidates; experiment
favors the $J = 2$ assignment.  In constructing Table \ref{tbl:delM},
we adjusted the 2P centroid to give the mass of the $J=2$ state in
agreement with observation.  At the observed mass, we compute
$\Gamma(2\slj{3}{2}{2} \to D\bar{D}) = 21.5\mev$ and
$\Gamma(2\slj{3}{2}{2} \to D\bar{D}^{*}) = 7.1\mev$, in reasonable
agreement with experiment (see Table \ref{tbl:decay}) We show in
Figure~\ref{fig:quattro} the
variation of the 2\slj{3}{2}{2} pole position with bare mass in 10-MeV 
steps, using Eqn.~(\ref{eq:redH}) to fully account for finite-width 
effects.

\subsubsection{$X(3872)$ as 2\slj{3}{2}{1} \label{subsubsec:x23p1}}
If $X(3872)$ is a $1^{++}$ state, the unique surviving $c\bar{c}$
candidate would be the 2\slj{3}{2}{1} level~\cite{Suzuki:2005ha}.  One
problem with this interpretation is clear:  the expected mass (cf.\
Table~\ref{tbl:delM}) is $48\mev$ above the observed $X$ mass.  If
$Z(3930)$ is indeed the 2\slj{3}{2}{2} level, it is very unlikely that
its $J=1$ partner should lie at $3872\mev$.  

If the mass were not a problem, the 2\slj{3}{2}{1} interpretation would
still be difficult to sustain.  The E1 radiative transition to $\gamma
\jpsi$ is greatly suppressed because of a cancellation between the
$c\bar c$ and $D\bar D^*$ contributions.  We obtain 35 and $21\kev$ for
the transitions to $\jpsi$ and $\psi^{\prime}$ respectively from the
$c\bar c$ contributions alone.  When the effects of the charmed meson
virtual channels are included, the rates change dramatically to 2.4 and
$90\kev$, respectively.  This would make the reported observation of
the $\gamma\jpsi$ mode highly unlikely.  Of course, within the \ccc\
model the isospin breaking between $D^0\bar D^{*0}$ and $D^+ D^{*-}$
virtual states is relatively small.  Even if we assume it to be large,
so that only the virtual $D^0\bar D^{*0}$ component is important, our analysis
suggests that a search for the radiative decay to $\psi^{\prime}$
would be valuable.

\begin{figure}[tb]
    \BoxedEPSF{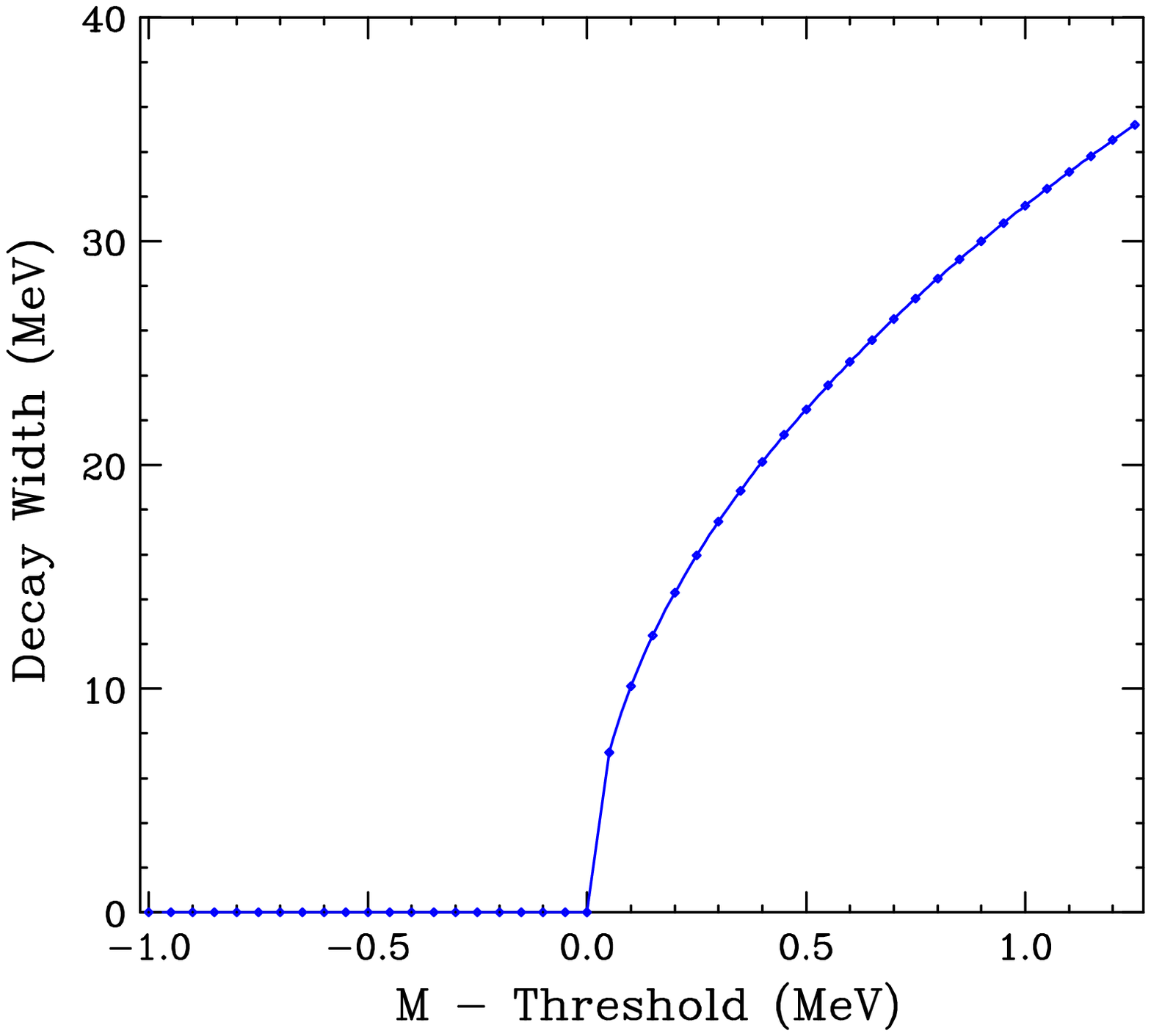 scaled 450}
    \caption{Mass dependence of  $\Gamma(2\slj{3}{2}{1}) \to D\bar D^*$.
    \label{fig:cinque}}
\end{figure}

Finally, the decay rate of  $2\slj{3}{2}{1} \to D\bar D^*$ increases very
rapidly above threshold as shown in Figure~\ref{fig:cinque}. 
Since the total width of $X(3872)$ is less than $2.3\mev$, the
possibility that the 2\slj{3}{2}{1} level lies more than $22\kev$ (!)
above the $D^{0}D^{*0}$ threshold is precluded.  When the additional
partial width $\Gamma_{\mathrm{other}}$ of the 2\slj{3}{2}{1} into
other modes ($ggg$, $\gamma\jpsi$, etc.)  is taken into account, the
largest tenable mass is lowered to $M(D^{0}) + M(D^{*0}) -
\Gamma_{\mathrm{other}}$.  Combining this result with the large ratio
of $\Gamma(X(3872) \to D^0 \pi^0 \bar D^0) \approx 10 \Gamma(X(3872)
\to \pi\pi \jpsi)$ reported by Belle~\cite{ShenLP05} renders the
2\slj{3}{2}{1} interpretation very improbable.

\subsection{$Y(4260)$ and other states \label{subsec:Babar}}
In initial-state radiation events, BaBar~\cite{Aubert:2005rm} has
recently reported a $\pi\pi\jpsi$ resonance, $Y(4260)$, with a width of
approximately $88\mev$.  This state is hard to assign within the
$c\bar c$ spectrum.  To be directly produced from a virtual photon, it
must be a $J^{PC}=1^{--}$ S or D state.  The small contribution to
$\Delta R$ discourages the 4S interpretation, but leaves open the
2\slj{3}{3}{1} identification.  However the \ccc\ analysis requires
that state at $4.16\gev$ in order to reproduce the structure of $\Delta
R$.  The identification of the $Y(4260)$ with the 2\slj{3}{3}{1} state is ruled
out in the \ccc\ model because the partial decay width of the 2D state
(at $4260\mev$) into charm meson pairs is $125\mev$.  This is not the
discovery mode for $Y(4260)$ but exceeds the reported total width.
Therefore, this state does not have a conventional charmonium
interpertation.  One plausible explanation for this state is that it is
a hybrid state~\cite{Zhu:2005hp,Close:2005iz}.  The lattice
calculations of Juge and collaborators~\cite{Juge:1999ie,Juge:2002br} for the $Q\bar Q$
potentials with excited gluonic modes give a lowest multiplet ($H_1$
in their notation) consisting of a quark spin-singlet $1^{--}$ state and
a spin-triplet set of states ($0^{-+}, 1^{-+},\hbox{ and }2^{-+}$).  These
states would be degenerate in the absence of relativistic
spin-dependent mass corrections.  If this interpretation were correct,
then analogous narrow states would be expected in the $b\bar b$ system.

\section{Summary \label{sec:sum}}
The computed properties of the  2\slj{3}{2}{2} and
3\slj{1}{1}{0} $c\bar c$ levels are in rough agreement with the $Z(3931)$ and
$X(3943)$ states observed at Belle.  We await more statistics and
confirmation of these results to make a  definitive statement on
the identification of these states.

It is extremely important to solidify what appears to be known about 
$X(3872)$, the best studied of the new states, beginning with its 
$1^{++}$ quantum numbers. We emphasize the need 
to confirm the $\gamma\jpsi$ decay mode, which fixes the $C=+$ 
assignment, and to determine the $X(3872) \to \pi^{0}\pi^{0}\jpsi$ 
branching fraction.

Experiments should continue to search for additional narrow charmonium
states in neutral combinations of charmed mesons and anticharmed
mesons.  The most likely candidates correspond to the $1\slj{3}{3}{3}$,
$1\slj{3}{3}{2}$, and $1\slj{3}{4}{4}$ levels~\cite{Eichten:2004uh}.
The analysis we have carried out can be extended to the $b\bar{b}$
system, where it may be possible to see discrete threshold-region
states as well.  For any of these new states that are threshold bound
states their mass is primarily determined by the positions of open
flavor channels.  For example, if the $X(3872)$ arises through
additional interaction in the two-meson sector, then in the $b\bar b$
system we would expect an analog state not far below $B^*\bar B$
threshold.

\begin{acknowledgments}
We thank our experimental colleagues for stimulating discussions.  KL's
research was supported by the Department of Energy under
Grant~No.~{DE--FG02--91ER40676}.  Fermilab is operated by
Universities Research Association Inc.\ under Contract No.\
DE-AC02-76CH03000 with the U.S.\ Department of Energy.

\end{acknowledgments}

\end{document}